\documentclass[twocolumn,aps,superscriptaddress,showpacs,preprintnumbers,amsmath,amssymb]{revtex4-1}
\usepackage{graphicx} 
\usepackage{bm} 

\newcommand{\bra}[1]{\langle #1 |}
\newcommand{\ket}[1]{| #1 \rangle}

\newcommand{\beq}{\begin{equation}}
\newcommand{\eeq}{\end{equation}}

\newcommand{\creation}[2]{\hat #1^{\dagger}_{#2}}
\newcommand{\annihilation}[2]{\hat #1_{#2}}

\def\vec#1{\mathbf{#1}}

\def\ri{\mathbf{r}_i}
\def\rj{\mathbf{r}_j}
\def\xi{\mathbf{x}_i}

\def\Ri{\mathbf{R}_i}

\begin{document}

\title{Tunable Holstein model with cold polar molecules}

\author{Felipe Herrera}
\email{fherrera@chem.ubc.ca}
\affiliation{Department of Chemistry, University of British Columbia, Vancouver, B.C., V6T 1Z1, Canada}

\author{Roman V. Krems}
\affiliation{Department of Chemistry, University of British Columbia, Vancouver, B.C., V6T 1Z1, Canada}
\date{\today}

\begin{abstract}

We show that an ensemble of polar molecules trapped in an optical lattice can be considered as a controllable open quantum system. The coupling between collective rotational excitations and the motion of the molecules in the lattice potential can be controlled by varying the strength and orientation of an external DC electric field as well as the intensity of the trapping laser. The system can be described by a generalized Holstein Hamiltonian with tunable parameters and can be used as a quantum simulator of excitation energy transfer and polaron phenomena. We show that the character of excitation energy transfer can be modified by tuning experimental parameters.   
\end{abstract}

\maketitle


{\it Introduction}.--
Many important biological processes involve energy transfer between complex molecules in mesoscopic aggregates \cite{Engel:2007,Scholes:2006}.  Energy can generally be transferred incoherently via direct two-molecule interactions or through the emergence of collective coherence described by excitons \cite{Agranovich:2008}. Incoherent energy transfer results from the interaction of excitons with phonons. It is a highly debated open question whether  exciton-phonon interactions conspire to ensure the most efficient and unidirectional energy transfer in biological systems \cite{Huelga:2008,Caruso:2009, Sarovar:2010}.  The exciton-phonon interactions can be described by a Holstein model \cite{Holstein:1959,Mahan}. It is difficult to calculate numerically the full energy spectrum for this polaron model in a complete range of interaction parameters \cite{Goodvin:2006}. Therefore, it is necessary to design an experimentally accessible many-body quantum system that would be described by the Holstein Hamiltonian with tunable parameters and arbitrary dimensionality. Such a system could be used for quantum simulation of excitation energy transfer (EET) in complex molecular aggregates and polaron phenomena in general.

There is currently growing interest in using ultracold atoms trapped on an optical lattice for quantum simulation of condensed matter physics \cite{Bloch:2005,Bloch:2008}. Ultracold atoms offer the possibility of designing systems that are well described by model Hamiltonians such as the Bose-Hubbard Hamiltonian \cite{Lewenstein:2007}. 
It was recently shown that polar molecules trapped on an optical lattice provide new possibilities for quantum simulation due to the presence of long-range dipole-dipole interactions \cite{Micheli:2006, Herrera:2010, Jesus:2010}. Here, we consider the interaction of rotational excitons with phonons in an ensemble of ultracold polar molecules trapped in an optical lattice. We show that, although the translational motion of molecules is largely determined by the intensity of the trapping laser, the dipole-dipole interaction can be made large enough to couple the dynamics of rotational excitons with the lattice vibrations. We demonstrate that by tuning the trapping laser intensity and an applied DC electric field, the strength of the exciton-phonon coupling can be controlled, and that the character of EET can be modified dynamically from coherent to incoherent.

{\it Molecular crystal Hamiltonian}.-- 
We consider an array of $^1\Sigma$ polar molecules in the rovibrational ground state, trapped in a 3D optical lattice with one molecule per lattice site and no tunneling of molecules between sites \cite{Bloch:2008,Lewenstein:2007}. Trapping of $^1\Sigma$ diatomic molecules on an optical lattice has recently been demonstrated \cite{Ospelkaus:2006,Danzl:2009, Lang:2008}. For the lowest  bands of the periodic lattice potential, molecules vibrate harmonically around the equilibrium positions $\Ri$ \cite{Bloch:2005,Lewenstein:2007}.

The trapping strength of the optical lattice is one experimental parameter to control the system. Another parameter can be introduced by applying a DC electric field $\mathbf{E}$. In a weak DC field, the rotational ground state $\ket{g}\approx a\ket{N=0,M_N=0}+b\ket{N=1,M_N=0}$ and the excited state $\ket{e}\approx b\ket{N=0,M_N=0}-a\ket{N=1,M_N=0}$ constitute an isolated two-level system \cite{Herrera:2010, Jesus:2010}. 
The field-free rotational states $\ket{N,M_N}$ are eigenstates of the rigid rotor Hamiltonian $\hat H_\text{R}=B_e\hat N^2$, where $B_e$ is the rotational constant.
The states $\ket{g}$ and $\ket{e}$ are eigenstates of the Hamiltonian $\hat H_\text{DC} =  \hat H_\text{R} - \mathbf{d}\cdot\mathbf{E}$, where $\mathbf{d}$ is the electric dipole operator. The coefficients $a$ and $b$ are functions of the DC field strength $E$.  
The electric dipole-dipole interaction $\hat V_{\text{I}}(\ri,\rj)$ couples the rotational states of molecules in different lattice sites.
For lattice site separations $a_L\approx$ 500 nm and molecules with a permanent dipole moment $d\ge 1$ Debye, the characteristic energy of the dipole-dipole interaction $V_{dd}\equiv d^2/a_L^3$ is a few tens of kHz.
In the two-molecule subspace $\mathcal{S}=\{\ket{g,g},\ket{g,e},\ket{e,g}\}$, the dipole-dipole operator $\hat V_{\text{I}}(\ri,\rj)$ has the following matrix elements: $V_{ij}^{gg}= \bra{g_i,g_j}\hat V_{\text{I}}\ket{g_i,g_j}$, $V_{ij}^{eg}= \bra{e_i,g_j}\hat V_{\text{I}}\ket{e_i,g_j}$, and $J_{ij} = \bra{g_i,e_j}\hat V_{\text{I}}\ket{e_i,g_j}$. These integrals can be evaluated as in Ref. \cite{Carr:2009}.  $V_{ij}^{gg}$ can be written as $V_{ij}^{gg}=U_g/|\ri-\rj|^3$. 

An ensemble of polar molecules in an optical lattice can be represented by a Hamiltonian of the form $\hat{\mathcal{H}}=\hat H_{\text{ph}}+\hat H_{\text{ex}}+\hat H_{\text{int}}$. The first term, 
\begin{equation}
 \hat H_{\text{ph}}= \sum_i^{} \frac{\vec{p}_i^2}{2m} + \sum_{i,j>i}V_{g}(\vec{r}_i,\vec{r}_j)\equiv\sum_{k,\nu}\hbar\omega_{k,\nu}\creation{a}{k,\nu}\annihilation{a}{k,\nu},
\label{phonon}
\end{equation}
describes phonons associated with the oscillatory motion of the molecules in the lattice potential $V_g(\ri,\rj)= m\omega_0^2(\vec{r}_i - \vec{R}_i)^2/2+ U_g/|\ri-\rj|^3$. The first contribution to $V_g$ depends on the intensity of the trapping laser that determines the trapping frequency $\omega_0$ \cite{Bloch:2005} and the molecular mass $m$. The second term in Eq. (\ref{phonon}) depends on the strength of the dipole-dipole interaction and couples the motion of molecules in different sites. The operator $\creation{a}{k,\nu}$ creates a phonon in mode $k$ with polarization $\nu=x,y,z$. A competition betwen the laser trapping force $f_L\propto m\omega_0^2$ and the dipole-dipole force $f_{dd}\propto U_g/a_L^5$ determines the phonon frequency $\omega_{k,\nu}$. 
For example, for a homogeneous 1D array $\omega_{k}=\omega_0\sqrt{1+12\rho\gamma(k)}$, where $\rho=(U_g/a_L^5)/ m\omega_0^2$ and $\gamma(k)=\sum_{j>0}[1-\cos(jk)]/j^5$, with $k$ within the first Brillouin zone. 
For finite arrays, the phonon frequencies are obtained from the eigenvalues of the force constant matrix ${\mathcal{F}}$ given by the Hessian of the lattice potential $V=\sum_{i,j} V_g(\ri,\rj)$ \cite{Wilsonbook}. The phonon spectrum in optical lattices is gapped for any value of $\rho$, i.e., $\omega_{k}\rightarrow \omega_0$ as $k\rightarrow0$, which resembles optical phonons in solids. In the limit $\rho\ll 1$, the spectrum is dispersionless (Einstein oscillators). This limit can be achieved either by increasing the lattice depth or by decreasing the DC field strength. In experiments, the Gaussian profile of the trapping beams usually generates an additional global harmonic potential $V_{\text{h}}(\ri)=m\omega_{\text{h}}^2\ri^2/2$. This potential can be included in $V_g(\ri,\rj)$. However, for the optical lattice potentials considered here $V_{\text{h}}$ can be neglected since $\omega_{\text{h}}/\omega_0<10^{-2}$ \cite{Bloch:2005}.



The second term in the total Hamiltonian $\hat{\mathcal{H}}$ describes collective rotational excitations (excitons) and is given by \cite{Agranovich:2008,Herrera:2010, Rabl:2007}
\begin{equation}
\hat H_{\text{ex}} = \sum_i (\epsilon_{eg}+D_i)\creation{B}{i}\annihilation{B}{i} + \sum_{i,j\neq i} J_{i,j}\creation{B}{i}\annihilation{B}{j}.
\label{exciton}
\end{equation}
In the present work, we neglect non-linear exciton-exciton interactions, which is a good approximation for a small number of rotational excitations. The transition operator $\hat B_i^{\dagger} = \ket{e_i}\bra{g_i}$ creates a rotational excitation  $\ket{g}\rightarrow \ket{e}$ in site {\it i}. The first term in Eq. (\ref{exciton}) contains the excitation energy at each site, which is a sum of the single-molecule rotational splitting $\epsilon_{eg}$ and the site-dependent shift $D_i=\sum_{j\neq i}D_{ij}$ due to the dipole-dipole interaction between molecules, where $D_{ij}=V_{ij}^{eg}-V_{ij}^{gg}$. The second term in Eq. (\ref{exciton}) describes the hopping of the rotational excitation between sites. The integrals $D_{ij}$ and $J_{ij}$ in Eq. (\ref{exciton}) are evaluated for molecules fixed at their equilibrium positions $\Ri$.

The final contribution to the system Hamiltonian $\hat{\mathcal{H}}$ is the exciton-phonon interaction. This effective coupling between internal and translational degrees of freedom of the molecules can be written as 
\begin{eqnarray}
\hat H_{\text{int}}^{\nu}&=&\sum_{k,i} g_{D_{i}}^{k\nu}\left(\creation{a}{k\nu}+\annihilation{a}{k\nu}\right)\hat B^{\dagger}_i\hat B_i \nonumber\\
&&\;\;\;\;\;\;\;\;\;\;+ \sum_k \sum_{i,j\neq i} g_{J_{ij}}^{k\nu}\left(\creation{a}{k\nu}+\annihilation{a}{k\nu}\right)\hat B^{\dagger}_i\hat B_j  ,
\label{interaction}
\end{eqnarray}
which derives from the Taylor expansion of the integrals $D_{ij}$ and $J_{ij}$ in Eq. (\ref{exciton}) up to linear order with respect to the small variation of the relative distance $\delta \vec{r}_{ij}$ from its equilibrium value, due to the motion of the molecules in their local potentials. The exciton-phonon coupling parameters $g_{D_{i}}^{k\nu}$ and $g_{J_{ij}}^{k\nu}$ can be obtained analytically from the gradient of the dipole-dipole potential and depend on the phonon mode $(k,\nu)$. For finite arrays, the mode dependence can be obtained from the eigenvectors of the force constant matrix $\mathcal{F}$. The first term in Eq. (\ref{interaction}) represents phonon-modulated site energies, proportional to $D_{ij}a_L^{-1}$, and the second term corresponds to phonon-modulated hopping of an excitation, proportional to $J_{ij}a_L^{-1}$.  Although the energy shift $D_{ij}$ is a small perturbation to the rotational spectrum, i.e., $D_{ij}\ll\epsilon_{eg}$, its fluctuation with the motion of the molecules in their local potentials can lead to dynamical localization of excitons \cite{Agranovich:2008, Caruso:2009}.
\begin{figure}[t]
\includegraphics[width=0.49\textwidth]{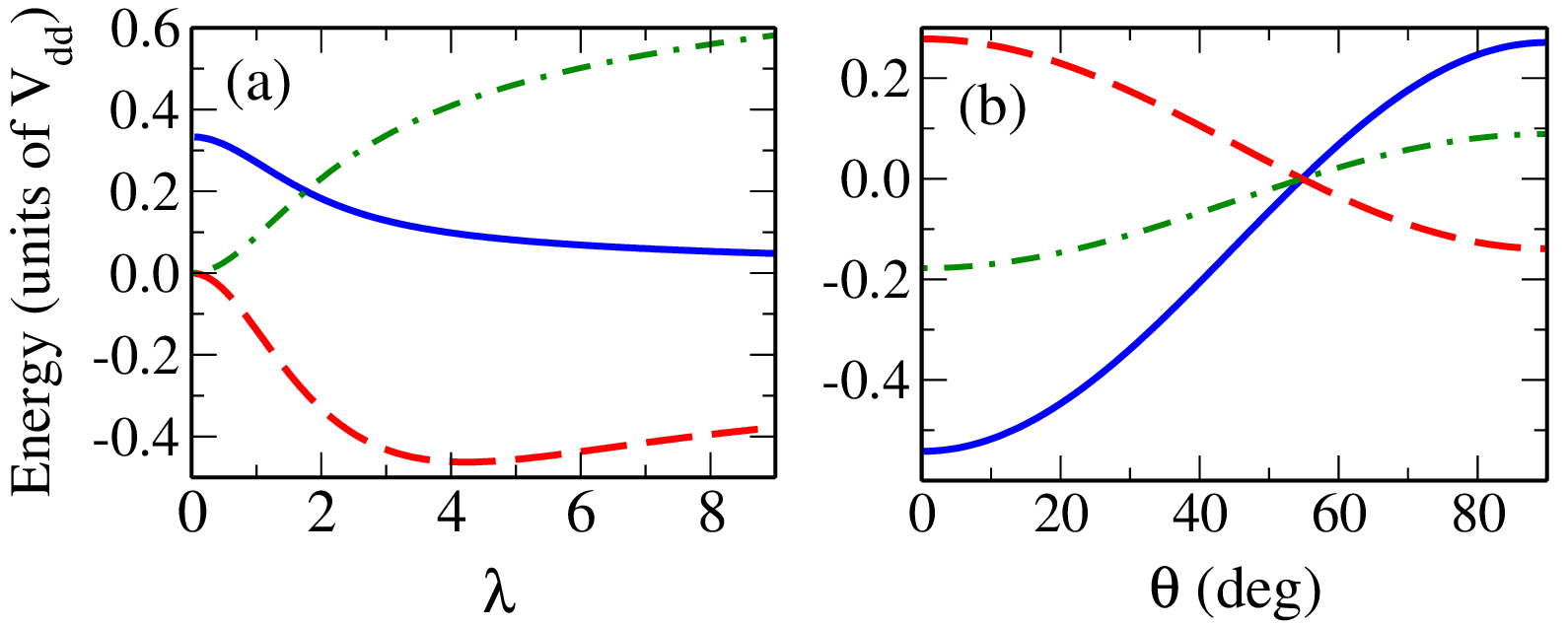}
\caption{(color online) Nearest-neighbour couplings $J_{12}$ (solid line), $D_{12}$ (dashed line), and $V_{12}^{gg}$ (dot-dashed line): (a) dependence on the DC field strength parameter $\lambda = dE/B_e$ for a DC field $\mathbf{E}=E\hat{\mathbf{z}}$ perpendicular to the one-dimensional array; (b) dependence on the angle $\theta$ between the electric field and the array, for $\lambda = 1$. Energy in units of $V_{dd} = d^2/a_L^3$, where $d$ is the permanent dipole moment and $a_L$ is the lattice constant. $B_e$ is the rotational constant of the molecule.}
\label{fig:Figure1}
\end{figure}

The system Hamiltonian $\hat{\mathcal{H}}$ in Eqs. (\ref{phonon})-(\ref{interaction}) constitutes a generalized polaron model. In the limit $g_{D_{i}}^{k\nu}\gg g_{J_{ij}}^{k\nu}$, it corresponds to the standard Holstein model \cite{Holstein:1959, Mahan}, extensively used to study energy transfer in molecular crystals and photosynthetic complexes \cite{Caruso:2009, Sarovar:2010}, polaron physics in solids \cite{Alexandrov:1995}, and, since recently, quantum information processing in dipolar gases \cite{Rabl:2007}. In the opposite limit $g_{D_{i}}^{k\nu}\ll g_{J_{ij}}^{k\nu}$, the Hamiltonian corresponds to the Su-Schrieffer-Heeger (SSH) model of particle-boson coupling, introduced to describe electrons in one-dimensional chains of polyacetylene \cite{Su:1979,Marchand:2010}. It is easy to include additional effects such as the anharmonicity of the optical lattice potential \cite{Zolotaryuk:1998}, quadratic exciton-phonon coupling \cite{Munn:1978} and exciton-exciton interactions \cite{Agranovich:2008} in the Hamiltonian. 

A similar polaron model $\hat{\mathcal{H}}$ was used in Ref. \cite{Rabl:2007} to describe 2D self-assembled crystals of polar molecules in the context of quantum information processing. Quasi-2D crystals are predicted to form at temperatures $T\sim 10$ nK, when the molecular kinetic energy is smaller than $V_{dd}$. A strong transverse confinement and a DC electric field perpendicular to the crystal plane are needed to stabilize the crystal against attractive dipole-dipole interactions. Self-assembled crystals exhibit acoustic phonons and the exciton-phonon interaction is always strong due to the presence of low frequency modes. These constraints limit the range of the Hamiltonian parameters $(J,D,g_{D}^{k},g_{J}^{k})$ that can be explored with such systems. In the system proposed here, molecules are stabilized against collisional losses by the optical lattice potential for any orientation of the DC electric field, and for any dimensionality. In addition, the coupling parameters can be tuned in a much wider range of values, as demonstrated below. 


{\it Tunable exciton-phonon coupling}.-- 
We now specialize our discussion to a finite 1D array of polar molecules. Effective lower-dimensional arrays in a 3D optical lattice can be generated when the dipole-dipole interaction is significant along one or two axes of the lattice only. We consider the interaction of a single exciton with harmonic phonons. 

The DC electric field modifies the single-molecule states $\ket{g}$ and $\ket{e}$, and therefore the value of the dipole-dipole couplings $J_{12}$, $V^{eg}_{12}$, and $V_{12}^{gg}$ \cite{Herrera:2010}. This is shown in Fig. \ref{fig:Figure1}(a) for a DC field perpendicular to the axis of the array. The magnitude of $J_{12}$ decreases with increasing field strength, whereas $D_{12}$ and $V^{gg}_{12}$ increase. For large DC fields, Eq. (\ref{interaction}) reduces to the Holstein polaron model plus a small correction due to the finite value of $g_{J_{12}}$. In the limit of weak DC fields, $D_{12}$ and $V_{12}^{gg}$ are vanishingly small due to parity selection rules, and Eq. (\ref{interaction}) reduces to the SSH polaron model with Einstein phonons.
\begin{figure}[t]
\includegraphics[width=0.49\textwidth]{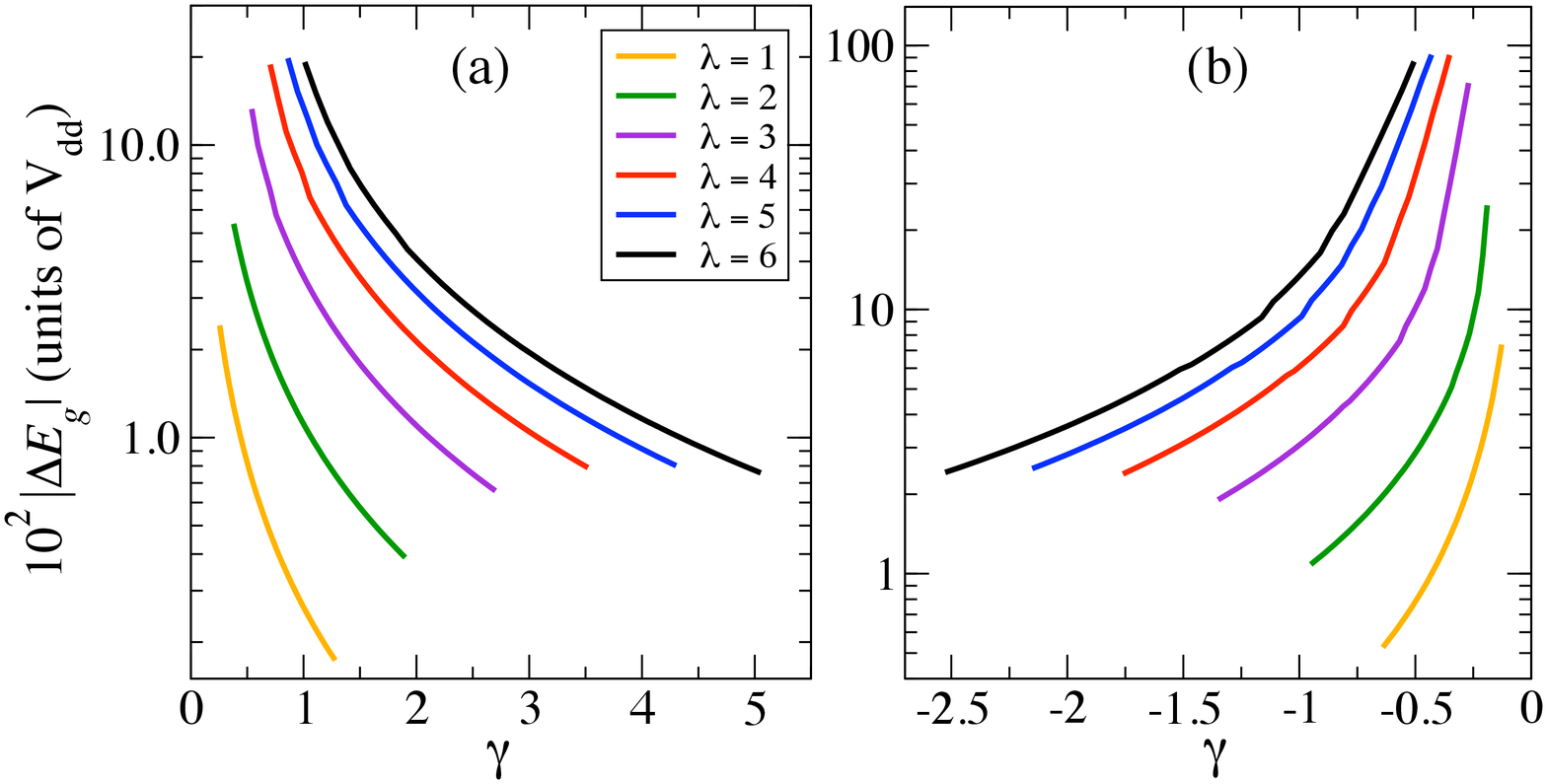}
\caption{(color online) Polaron shift $\Delta E_g$ as a function of the trapping frequency $\omega_0$ for an array of 10 LiCs molecules separated by 400 nm. Curves are shown for electric fields of 9 kV/cm (solid line), 2 kV/cm (dashed line), and 0.6 kV/cm (dotted line). Panels (a) and (b) correspond to a field perpendicular and parallel to the array, respectively. In panel (a) we have $|\Delta E_{g}|>J_{12}=6.73$ kHz, for $E=9$ kV/cm and $\omega_0/2\pi < 20$ kHz, which is a signature of strong exciton-phonon coupling.}
\label{fig:polaron}
\end{figure}
The values of the dipole-dipole matrix elements also depend on the angle $\theta$ between the DC electric field and the molecular array. For our chosen rotational subspace with projection $M_N=0$ along the electric field axis, the parameters $D_{12}$, $J_{12}$, and $V_{12}^{gg}$ are proportional to $(3\cos^2{\theta} -1)$ and vanish for $\theta \approx 54.7 ^{\circ}$, as shown in Fig. \ref{fig:Figure1}(b). 

The strength of the exciton-phonon coupling can be controlled by tuning the trapping laser intensity and the DC electric field. The coupling constants in Eq. (\ref{interaction}) can be written as $g_{\Lambda ij}\propto \sqrt{(1/m\omega_0)}(\Lambda_{12}/a)$, for $\Lambda=D,J$. 
In order to quantify the strength of this coupling for realistic trapping conditions, we analyze the eigenvalues of the Holstein Hamiltonian for a given molecular species in a finite 1D array. We diagonalize the total Hamiltonian $\hat{\mathcal{H}}$ numerically for an array of $\sigma$ molecules in the site basis $\ket{g_1,\ldots,e_i,\ldots,g_{\sigma}}\ket{\nu_1,\nu_2,\ldots,\nu_{\sigma}}$, where $\nu_k$ is the occupation number of the phonon mode $k$. The phonon basis is truncated by including states with up to a given phonon occupation $\nu_{\text{max}}$. The value of $\nu_{\text{max}}$ is increased iteratively until the calculated observable is converged.
We partition the Hamiltonian as $\hat{ \mathcal{H}} = \hat H_0 + \hat H_{\text{int}}$, where $\hat H_0 =\hat H_{\text{ex}}+\hat{H}_{\text{ph}}$. The ground state energy $E_g$ of the non-interacting Hamiltonian $\hat H_0$ is chosen as a reference. Any interaction between rotational excitons and the lattice vibrations of the molecules shifts the ground state $E_g$ towards lower energies. 
As an illustrative example, we show in Fig. \ref{fig:polaron} the shift $\Delta E_g$ for a finite array of LiCs molecules ($d = 5.5$ Debye \cite{Deiglmayr:2008}) separated by $a_L=400$ nm, as a function of the optical lattice trap frequency $\omega_0$.  Each curve corresponds to a different field strength $E< 10$ kV/cm. For small trapping frequencies (weaker lattices), we find $\Delta E_g\geq J_{12}$, which is a signature of strong coupling \cite{Mahan}. The strength of the exciton-phonon coupling is larger for a DC field parallel to the array than for any other field orientation. 

{\it Application to energy transfer in a phonon bath.--}
One of the possible applications of a tunable Holstein Hamiltonian using cold polar molecules is the simulation of excitation energy transfer processes (EET) that occur in molecular crystals and light-harvesting complexes at room temperature \cite{Scholes:2006, Engel:2007,Caruso:2009, Sarovar:2010, Mukamel:2000}. For example, let us consider two LiCs molecules in an optical lattice with $a_L=400$ nm and trapping frequency $\omega_0/2\pi =10$ kHz, in a DC field of 10 kV/cm perpendicular to the intermolecular axis. The two normal modes of lattice vibration have frequencies $\omega_1=\omega_0$ and $\omega_2=2.8\omega_0$. The lower frequency mode does not couple to excitons because it does not change the relative distance between molecules. The Hamiltonian parameters for the higher frequency mode are $g_{D_{12}}/h=12.7$ kHz and $g_{J_{12}}/h=-2.33$ kHz. This gives the ratio $g_{D_{12}}/J_{12}\approx 2$, which can also be found in the Fenna-Mathews-Olson photosynthetic complex, where electronic excitations are believed to be locally coupled to phonons at each site \cite{Sarovar:2010, Caruso:2009}. For a weak DC field $E=$ 0.5 kV/cm, the first term in Eq. (\ref{interaction}) is negligible, and non-diagonal coupling dominates ($g_{J_{12}}/J_{12}\approx 0.6$, for $\omega_0/2\pi=10$ kHz), which may allow for tests of the role of spatial non-local phonon correlations in the dynamics of EET \cite{Meier:1997}.

In order to model rotational EET using the generalized polaron Hamiltonian in Eqs. (\ref{phonon})-(\ref{interaction}), we define an initial wavefunction $\ket{\Psi(0)}$ describing the coupled exciton-phonon system, using the product basis described above. We propagate the time-dependent Schr\"{o}dinger equation, using the Hamiltonian $\hat{\mathcal{H}}$, and construct the total density matrix $\hat\rho(t)=\ket{\Psi(t)}\bra{\Psi(t)}$ at each time step. We then obtain the reduced density matrix in the exciton subspace $\hat\rho_{E}(t) = \text{Tr}_{\text{vib}} \{\hat\rho_{}(t)\}$ by tracing over the states in the truncated phonon basis. The diagonal elements of the reduced density matrix are the time-dependent probabilities $p_i(t)$ for a molecule in site {\it i} to be in the rotational excited state $\ket{e_i}$. 
\begin{figure}[t]
\includegraphics[width=0.49\textwidth]{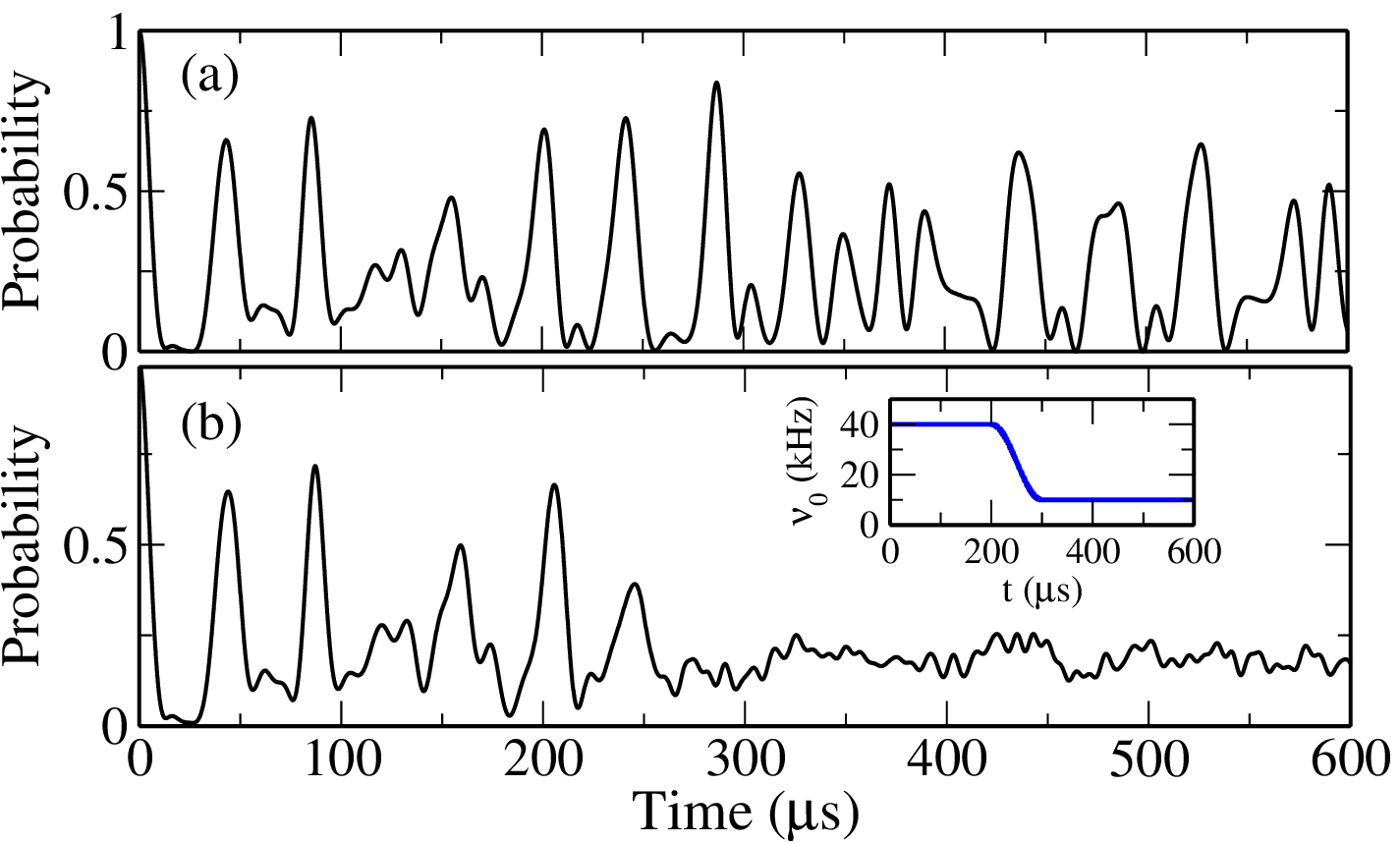}
\caption{(color online) Excitation energy transfer in an array of five LiCs molecules in a DC electric field perpendicular to the array: (a) Evolution of the excitation probability $p_1(t)$, when no phonons are present; (b) The same as in (a), but with phonons in an optical lattice with trapping frequency $\nu_0=\omega_0/2\pi$ varying in time as indicated in the inset. The field strength is 0.5 kV/cm.
}
\label{fig:transfer}
\end{figure}
We apply this procedure to a 1D array of five LiCs molecules separated by 400 nm, in a weak DC field $E = 0.5$ kV/cm perpendicular to the array. The rotational excitation is initially in molecule $1$. Single-site excitation and probing of rotational states can be achieved by applying an electric field gradient as described in Ref.  \cite{DeMille:2002}. The phonon bath has initially zero temperature. In Fig. \ref{fig:transfer}, we show the dynamics of the excitation at site 1 for different trapping frequencies $ \omega_0/2\pi$, keeping $E$ constant. 
In Fig. \ref{fig:transfer}(a) we set $\omega_0\rightarrow \infty$, thereby neglecting the coupling to phonons. In this limit, the excitation energy oscillates coherently between the molecules of the array \cite{Engel:2007}, with a transfer time between adjacent sites $\tau_{12} = h/|J_{12}|\sim 10 \;\mu s$. 
When the exciton-phonon coupling is turned on, by decreasing the trapping frequency in this case, the transport of the excitation is suppressed due to a competition between the exciton-phonon interactions and the excitonic energy transfer. When the exciton-phonon coupling is strong enough, all the site populations $p_i(t)$ approach an equilibrium value. This is shown in Fig. \ref{fig:transfer}(b), where only the population in site 1 is presented for simplicity. We obtain a similar behaviour when $\omega_0$ is kept constant and the DC field strength is dynamically tuned from weak to strong.

{\it Discussion}.-
We have shown that the translational and rotational states of polar molecules trapped in an optical lattice can interact in the presence of a DC electric field. This interaction is described by a generalized polaron model with tunable parameters. All the terms in the Hamiltonian can be dynamically tuned by varying the DC electric field or the lattice laser intensity. 

These results suggest the possibility of using cold polar molecules for quantum simulation of excitation energy transfer processes that occur at high temperatures in solids and mesoscopic systems of biological interest \cite{Engel:2007,Scholes:2006, Agranovich:2008}. Despite the physico-chemical differences between these systems and cold polar molecules, they are described by the same Hamiltonian, and it should be possible to use the latter to experimentally study the role of environmental noise in the efficiency of quantum transport \cite{Caruso:2009,Rebentrost:2009,Segal:2007}. 
In addition, polar molecules can also be used to explore the strong coupling regime of exciton-phonon interaction, which may provide insight into some of the open questions of polaron physics, such as the role of the lattice dynamics in the mechanism of high-$T_c$ superconductivity \cite{Alexandrov:1995}.   

The interactions of atomic spin states with phonons are used in experiments with ultracold trapped ions to generate phonon-mediated gates for quantum computing \cite{Soderberg:2010}. Similarly, the interactions of rotational states with lattice vibrations described in this work can be used to produce novel quantum gates. The detrimental effects of decoherence can be minimized by increasing the trapping frequency. Phonon-mediated interactions may also introduce new adjustable parameters to the spin-lattice Hamiltonians that can be simulated with ultracold molecules \cite{Micheli:2006, Ortner:2009}. In particular, the controllable coupling to phonons can be exploited to study the effects of dynamically tunable decoherence on topologically ordered states that can be created with ultracold molecules \cite{Micheli:2006}. 
In summary, the proposed system may provide an experimental tool to explore the dynamics of quantum networks whose essential features are replicated in several areas of condensed matter physics. The ability to tune the system-environment coupling dynamically may also stimulate new studies of the dynamics of open quantum systems.  

{\it Note}.- After submission of this work, quantum simulation of EET using superconducting qubits was proposed in Ref. \cite{Mostame:2011}.

We thank Mona Berciu, Peter Rabl, Fillipo Caruso, and Martin Plenio for useful discussions. This work is supported by NSERC of Canada and the Peter Wall Institute for Advanced Studies. 

\bibliography{holstein}
\end{document}